\documentclass[]{spie}

\usepackage[]{graphicx}
\usepackage{amsmath}
\usepackage{amssymb}
\usepackage{floatflt}

\newcommand{\bnu}{\ensuremath{\boldsymbol{\nu}}}
\newcommand{\brho}{\ensuremath{\boldsymbol{\rho}}}
\newcommand{\br}{\ensuremath{\mathbf{r}}}
\newcommand{\bu}{\ensuremath{\mathbf{u}}}
\newcommand{\cns}{\ensuremath{C_{N}^{2}}}
\newcommand{\uij}{\ensuremath{U_{i,j}}}

\title{Adaptive optics point spread function reconstruction: \\lessons learned from on-sky experiment on Altair/Gemini and pathway for future systems}

\author{Laurent Jolissaint\supit{a}, Julian Christou\supit{b}, Peter Wizinowich\supit{c} and Eline Tolstoy\supit{d}
\skiplinehalf
\supit{a}aquilAOptics, 4, rue de la Ch\^{a}tellenie, 1635 La Tour-de-Tr\^{e}me, Switzerland; \\
\supit{b}Gemini Observatory, Hilo, Hawaii, USA;\\
\supit{c}W. M. Keck Observatory, Hilo, Hawaii, USA;\\
\supit{d}Kapteyn Astronomical Institute, Groningen, The Netherlands}
\authorinfo{Further author information: e-mail: laurent.jolissaint@aquilaoptics.com, web page: www.aquilaoptics.com}

\begin{document} 
\maketitle 

\begin{abstract}
We present the results of an on-sky point spread function reconstruction (PSF-R) experiment for the Gemini North telescope adaptive optics system, Altair, in the simplest mode, bright on-axis natural guise star. We demonstrate that our PSF-R method does work for system performance diagnostic but suffers from hidden telescope and system aberrations that are not accounted for in the model, making the reconstruction unsuccessful for Altair, for now. We discuss the probable origin of the discrepancy. In the last section, we propose alternative PSF-R methods for future multiple natural and laser guide stars systems.
\end{abstract}

\keywords{adaptive optics, point spread function reconstruction, calibration, telemetry}

\section{INTRODUCTION}\label{sec:intro}

Knowing the point spread function (PSF) associated to a given AO run is required for proper AO data reduction. Most of the time, there are no bright, well isolated star at an appropriate location in the science field of view that could be used as a clean reference PSF. An intuitive technique would be to image an AO-corrected isolated star right before and/or after the science AO run, but (1) seeing conditions vary rapidly so this PSF would not necessarily represents the conditions of the science run, (2) this would take precious telescope time to observe a point source. An alternative method was therefore proposed by V\'{e}ran et al. \cite{veran:97}, based on a simple fact: the residual wavefront all along the AO run is seen by the wavefront sensor (WFS), therefore it must be possible to reconstruct the long exposure PSF associated to the AO run from the WFS measurements. This technique was developed and successfully tested by V\'{e}ran et al. on PUEO, a curvature sensing AO system on the Canada-France-Hawaii telescope.

We later adapted the technique -- essentially the WFS noise determination method -- for a Shack-Hartmann based system (Jolissaint et al.\cite{jolissaint:04}), and applied it successfully to Altair, the Gemini-North telescope AO system, during Altair's commissioning in the laboratory. We only did a single PSF-R test at that time, fortunately this was enough to demonstrate that the algorithm was working well and produced a PSF that was close enough to the real PSF measured on the commissioning imager, knowing that the test conditions were not representative of the actual on-sky conditions (100 m/s wind, vibrations on the imager). Unfortunately, we never got the resources to test our PSF-R tool (OPERA) on real sky data, and as a consequence, this code was never made available for Altair's users community.

Since V\'{e}ran's success on PUEO, several teams have tried to implement the method for a variety of AO systems, some with success -- see a review in Le Mignant et al.\cite{lemignant:08}, but most failed to provide a finalized tool for usage by the AO astronomer. The reasons for failure were not technical but had to do with project management. The main issues we have identified amongst the different project were:\newpage
\begin{itemize}
\item partial access to the telemetry data needed to rebuild the PSF;
\item no access -- or not enough access -- to telescope engineering time;
\item bad communication between the engineer in charge of the PSF-R project and the people in charge of running the existing AO system, i.e. a lack of understanding of the AO operators for the needs in telemetry data;
\item impossible, or very difficult, to modify the control software of existing AO systems to gather telemetry data;
\item lack of human resources to finalize the PSF-R algorithm implementation in the instrument data pipeline;
\item lack of understanding of the AO user for the PSF-R algorithm procedure.
\end{itemize}
The method developed by V\'{e}ran was never demonstrated to be at the origin of the problem. A few adjustments were simply made to account for system particularities. One of the important consequence of these failures has been that PSF-R has been considered, in the AO users community, as something difficult or even impossible to do. We believe this situation has no scientific justification. Besides, PSF-R is requested by the AO users community more than never, as AO system becomes more and more complex, diversified, and popular (fitting a Moffat function to the AO PSF becomes less and less an acceptable solution because the PSF has a complex structure and varies across the field). Therefore, it is important now to finalize once for all a PSF-R algorithm as a service to the community, to demonstrate with sky data what we already know from the lab -- i.e. that V\'{e}ran's method does work for SH systems as well. From this, the AO engineers would get feed-back from AO astronomers on the usefulness and the practical limits of PSF reconstruction. Besides, developing a PSF-R tool is nowadays a specific requirement for future and near future AO systems (NFIRAOS on TMT, MUSE on the ESO's AO facility).

We have embarked ourselves in such a program, following a pragmatic, step-by-step approach. We have found two existing AO systems for which the limitations listed above are non existent, and where full telemetry has already been implemented in the past and are functional. These systems are Altair, the Gemini North AO system, and the AO system of Keck II telescope. Our objective is to implement on these systems the algorithm we wrote a few years earlier for Altair, OPERA, that was tested in the laboratory (a recall of our lab results is given below), without any new developments, as soon as possible, for the most simplistic AO mode: on-axis, bright NGS. The accent is put on code availability, not theoretical developments. Once the algorithm is implemented and functional, then we will start proposing improvements for (1) off-axis PSF-R, (2) dim NGS, (3) LGS systems, etc.

The rest of the paper is organized this way: first we review V\'{e}ran's method and our preliminary results on Altair lab tests, then we present and discuss our first on-sky results for Altair. At the time of the writing of this paper, OPERA was not yet available for Gemini-N user's community. We explain the issues we are facing right now with the system and propose solutions to solve these out. Based on our experience with a real system, we propose news ways to deal with system's hidden wavefront errors in the context of PSF-R.

\section{V\'ERAN METHOD FOR POINT SPREAD FUNCTION RECONSTRUCTION}

In a closed loop system, after the deformable mirror (DM) correction, the partially corrected beam is split between the science optical path and the WFS path. The WFS therefore sees, in principle, the same aberration than the science instrument, therefore by keeping all the WFS measurements it must be possible to reconstruct the PSF associated to the AO run. The most intuitive way of doing this would be to compute and sum all the intermediate instantaneous PSF associated to each WFS measurement. Unfortunately, there are plenty of reasons that make this approach difficult, eventually impossible.

First, the WFS sees only the aberrations at spatial frequencies below the AO cutoff frequency $1/(\text{2 WFS pitch})$: the non-corrected turbulent aberrations at high order are simply not seen by the system, and generate what is called the fitting error. Tis error is generally the main wavefront error component in classical AO systems, and not taking it into account would generate a very significant difference between the reconstructed PSF and the actual PSF. This would be already a good reason to drop the instantaneous PSF technique. But suppose that the WFS pitch is small enough so that high order aberrations are well sampled. Then we would have to deal with the WFS noise, which, having a white spectrum, add a constant spurious background on the reconstructed PSF, and the temporal error, due to the non-zero WFS integration time, which would smooth out the high spatial frequencies in the instantaneous aberration. In short, the instantaneous PSF procedure would produce a PSF with unrealistic wings structure and background (see Jolissaint et al.\cite{jolissaint:06} for a discussion on the impact of the classical AO errors on the PSF structure), and of course, a wrong Strehl ratio. And considering the DM commands instead of the WFS measurements improves nothing: the high order aberrations would remain unknown.

V\'{e}ran's method solve these issues by proposing a model that links the long exposure OTF and the covariances of the system's modes residual coefficients. In this approach, the different components of the WFS error measurement identified can be clearly identified and separated, allowing a "clean" determination of the covariance matrix of the true residual aberrations at the output of the system. Therefore, this method allows the reconstruction of the actual long exposure OTF/PSF. We will now give the summary of this method -- see reference papers given above for details.

The overall OTF of the system telescope + instrument + residual AO corrected turbulence can be written (this is an approximation) as the product of the telescope+instrument OTF and an equivalent AO OTF filter
\begin{equation}
\text{OTF}_{\text{system}}(\bnu)\approx\text{OTF}_{\text{telescope}}(\bnu)\,
\text{OTF}_{\text{AO}}(\bnu)
\end{equation}
The AO-OTF filter is directly associated to the spatial structure function of the residual phase in the telescope pupil, $D_{\varphi}$, following
\begin{equation}
\text{OTF}_{\text{AO}}(\bnu)=\exp{[-1/2\,\overline{D}_{\varphi}(\lambda\bnu)]}
\end{equation}
where $\lambda$ is the imaging wavelength. The structure function, a measure of the average quadratic difference between the phase taken at two different points separated by a vector \brho\ in the pupil, actually depends on where it is measured in the pupil, i.e. if \br\ represents the location in the pupil, and $\langle\cdot\rangle$ the ensemble average over an infinite number of realization of the turbulent phase,
\begin{equation}
D_{\varphi}(\brho,\br)=\langle[\varphi(\br+\brho,t)-\varphi(\br,t)]^{2}\rangle_{t}
\end{equation}
Indeed, the corrected phase is not stationary in the pupil, basically its RMS value increases from the center to the pupil edge. In order to make possible the convenient separation between the telescope and AO OTF, though, which simplify the model and makes the computation much easier, we use a structure function averaged over the pupil location, which s equivalent to assume that the phase is stationary over the pupil,
\begin{equation}
\overline{D}_{\varphi}(\brho)=\langle D_{\varphi}(\brho,\br) \rangle_{\br}
\end{equation}
This stationary assumption has a minimal impact. It only affects the low order aberrations which are well corrected in an AO system anyway, and the strong turbulence case, where the AO system does not performs well anyway.

V\'{e}ran's method makes use of a modal approach. This allows a convenient separation of the space vector the turbulent phase is associated to into two independent sub-spaces, first the space generated by the DM modes, and the complement of this mirror space, i.e. the space generated by the aberrations the DM cannot correct (generating the fitting error). Using a modal representation, the average structure function of the phase can therefore be split into a DM structure function and a fitting error structure function,
\begin{equation}
\overline{D}_{\varphi}(\brho)=
\overline{D}_{\varphi,\text{\tiny DM}}(\brho)+\overline{D}_{\varphi,\text{\tiny FE}}(\brho)
\end{equation}
Developing the DM structure function, one gets
\begin{equation}
\overline{D}_{\varphi,\text{\tiny DM}}(\brho)=\sum_{i,j}\langle\epsilon_{i}\epsilon_{j}\rangle\,\uij(\brho)
\end{equation}
where $\langle\epsilon_{i}\epsilon_{j}\rangle$ is the covariance of the residual AO corrected DM modes coefficients $\epsilon_{i}$ and $\epsilon_{j}$ (the residual coefficients hereafter), averaged over the N loop samples along the AO run,
\begin{equation}
\langle\epsilon_{i}\epsilon_{j}\rangle=
\frac{1}{N}\sum_{k=1}^{N}\,\epsilon_{i}(t_k)\,\epsilon_{j}(t_k)
\end{equation}
and $U_{i,j}$ is a sort of spatial covariance of the DM modes $M_{i}$ and $M_{j}$ defined by
\begin{equation}
\uij(\brho)=\frac{\iint_{\mathbb{R}^{2}}\,P(\bu)\,P(\bu+\brho)
\left[M_{i}(\bu)-M_{i}(\bu+\brho)\right]
\left[M_{j}(\bu)-M_{j}(\bu+\brho)\right]\,d^{2}u}
{\iint_{\mathbb{R}^{2}}\,P(\bu)\,P(\bu+\brho)\,d^{2}u}
\label{eq:uij}
\end{equation}
where $P$ is the pupil transmission (1 inside, 0 outside). The \uij\ functions can be computed once for all from the DM modes (in principle given numerically) and stored for use when needed. The residual coefficients covariance matrix is computed from the noise-cleaned WFS measurements covariance matrix and the command matrix $C$ (the one to go from the WFS measurements to the DM modes coefficients)
\begin{equation}
\langle\epsilon_{i}\epsilon_{j}\rangle=C\,\langle\omega_{i}\omega_{j}\rangle\,C^{t}+
\langle a_{i} a_{j}\rangle
\end{equation}
The noise free WFS measurements covariance $\langle\omega_{i}\omega_{j}\rangle$ are obtained from the noisy WFS measurements covariances from which an estimate of the noise has been subtracted out (the noise only affects the diagonal of the WFS covariance matrix). Good models for the WFS noise have been developed by V\'eran for a curvature sensing WFS and we have developed the same type of model for a SH-WFS. Details of the method are not recalled here for space reasons (see reference papers). $\langle a_{i} a_{j}\rangle$ is the covariance of the coefficients of the projection of the aliased phase onto the DM modes. As high order modes are not affected by AO correction, and as their amplitude only scales with the ratio $(D/r_0)^{5/6}$, the aliasing covariance matrix can be computed once for all from the modes, for a ratio $D/r_0=1$, and scaled once the actual $r_0$ associated with the AO run is known (this is explained below).

The fitting error structure function $\overline{D}_{\varphi,\text{\tiny FE}}$, as well, does not depend on the AO correction efficiency, and can be computed using different approaches, again for $D/r_0=1$, and scaled appropriately later. One approach (V\'eran's) makes use of a Monte-Carlo modeling of the high order phase and computes numerically the average fitting error structure function from this numerical experiment. An other approach (ours) consists in using the relationship between the phase spatial power spectrum and the phase structure function\cite{jolissaint:06}, which allows a direct and faster modeling of this FE structure function.

Fried parameter $r_0$ is obtained from the following procedure: as we know, in closed loop operation, the DM commands are set to make the DM surface following the incoming turbulent wavefront as close as possible. Therefore, the statistics of the DM modal coefficients should closely follow the statistics of the turbulent phase projected onto the DM modes basis. This later statistics is computed from the Von Karman (VK) model for the turbulent phase spatial power spectrum, projected onto the DM basis. By fitting the DM modal commands variances to the theoretical VK based model, one can extract the $r_0$ and outer scale $L_0$ associated with the AO run. The DM modal commands need to be cleaned up from command noise, and it is recommended to avoid using modes that are prone to non-turbulent perturbation, in particular tilt and possibly the second order modes (defocus and astigmatisms).

Now we are left with the determination of the telescope \& instrument OTF. The procedure proposed by V\'era, for this, is simple: as this OTF is not supposed to change with atmospheric conditions (but only with telescope orientation and gravity vector) the idea is to image a bright single isolated guide star and reconstruct the AO-OTF associated with it. By dividing the on-sky measured OTF by the AO-OTF, one gets the telescope OTF. This procedure needs to be done whenever it is felt that the telescope aberrations might have changed due to any change in the telescope operation, and should not take more than five minutes of telescope time overall (including one minute exposure to average out the speckle pattern).

\section{OPERA CODE FOR ALTAIR: PREVIOUS LABORATORY RESULTS}

Before going to the Gemini-North telescope, Altair was tested in the laboratory in real conditions (temperature, gravity vector, optical turbulence). We took this opportunity to test our PSF-R code OPERA at this time, and compare the predicted PSF with the PSF measured on the commissioning imager. The first important step was to test the seeing ($r_0,L_0$) estimator. We found that (1) the distribution of the mirror modes commands variances was following very well the expected Von-Karman based distribution, (2) the estimated $r_0$ was within the expected value given for the phase screens used to emulate optical turbulence.

The second step was to compute the PSF, and we found that the agreement with the actual PSF imaged on NIRI, the imager used for the experiment, was reasonably good: we predicted a Strehl of 0.125, and the measured Strehl was 0.15. The Strehl estimation on the real PSF image was approximate, though, as it depends sensibly on the estimation of the true image background. Normalized to their peak values, the PSF profiles were in excellent accord, though. See reference paper for details. To conclude, this first and only experiment in almost real conditions was demonstrating that the PSF-R method was working well, on an AO system that was working, in the lab, as expected. Additional tests would have been necessary to better understand the limits of the PSF-R technique, but these were never done.

\section{OPERA CODE FOR ALTAIR: FIRST ON-SKY RESULTS}

\subsection{Seeing estimation}

\begin{figure}[ht]
   \centering
   \includegraphics[width=8cm]{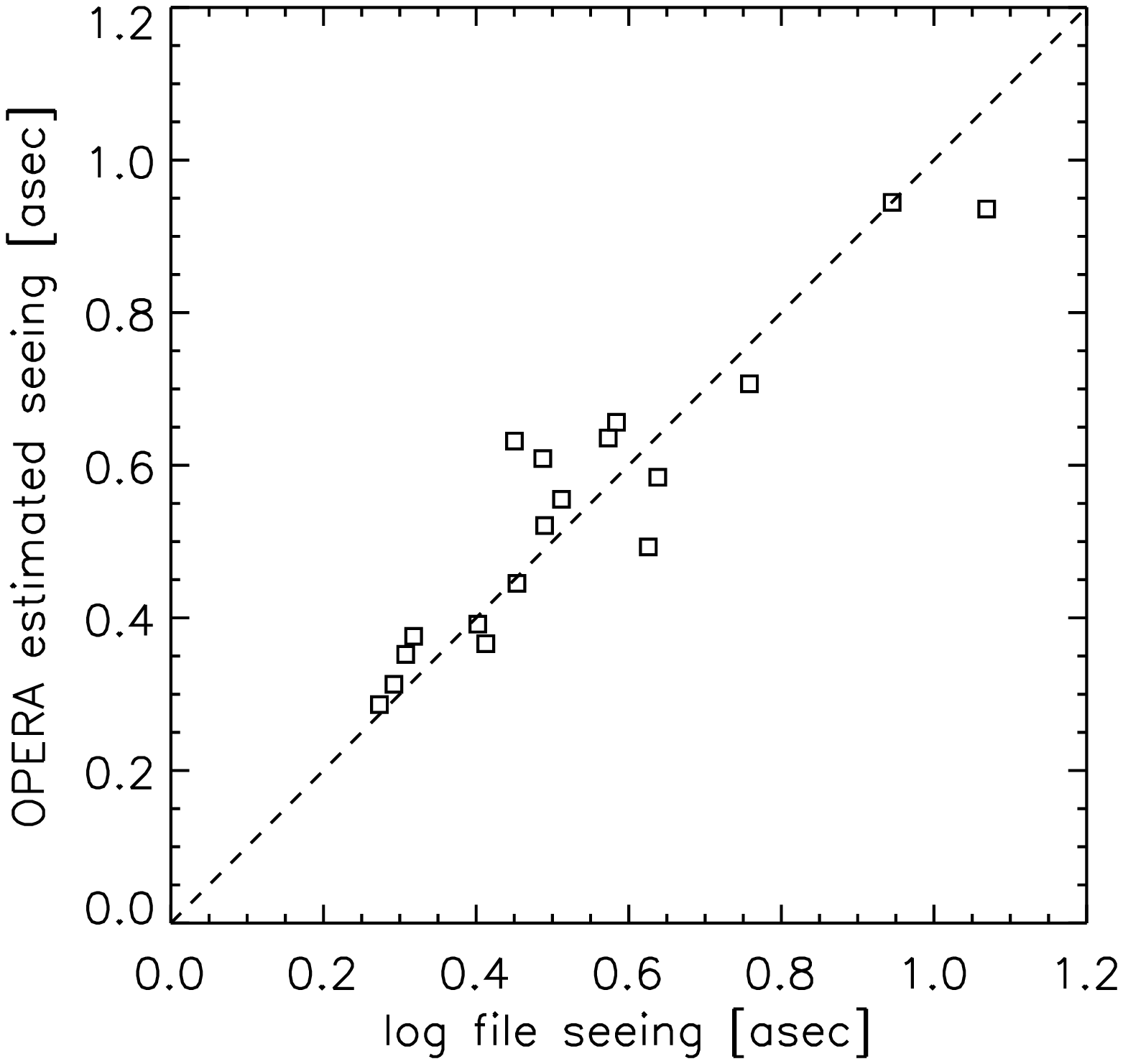}
   \includegraphics[width=8cm]{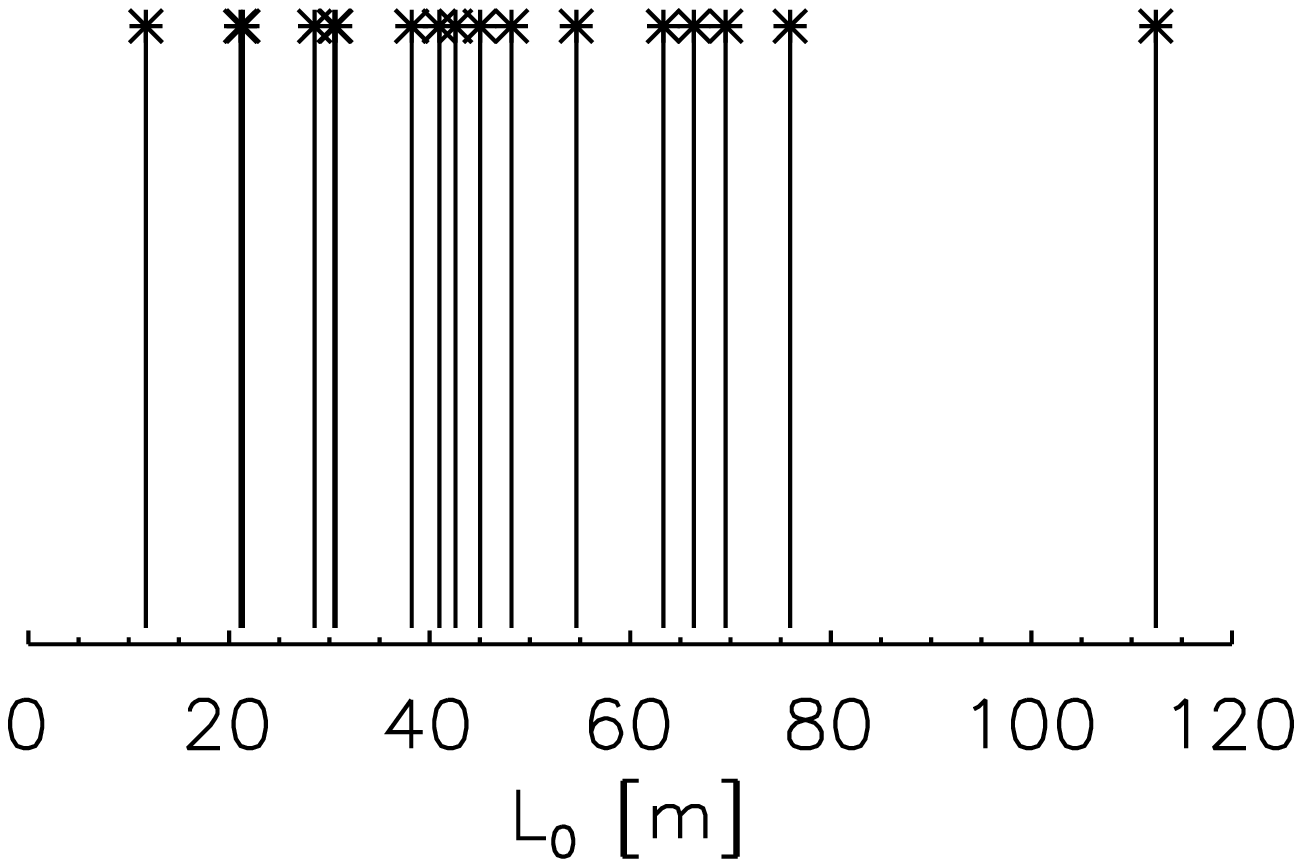}
   \caption{Left: seeing values retrieved from the fit of the modal commands to the Von Karman based model, as compared to the Altair seeing estimate log file, median seeing is 0.5". Right: outer scale values for the same nights, median value is 42 m.}
   \label{fig:fig01}
\end{figure}

We got 20 nights of Altair telemetry data for the month of December 2009. The NGS were bright (mV$<$8) and on-axis, so WFS noise could be neglected. To start, we estimated the seeing from the modal commands to the DM. Several comments can be made: first, the typical values found for the seeing (0.3" to 1", median 0.5") and outer scale (12 m to 120 m, median 42 m) were in good agreement with the typical Mauna Kea summit values. See figure \ref{fig:fig01}. The seeing results were also compatible with the values found in Altair's seeing log file for the same nights, even though it is not clear at this point if these values are comparable, as they were estimated using the very same tool, but in two different versions. In any case, a DIMM/MASS seeing monitor is now operational at the MK summit and we will make use of this tool in future comparisons with our own Altair based estimate.

\begin{figure}[ht]
   \centering
   \includegraphics[height=8cm]{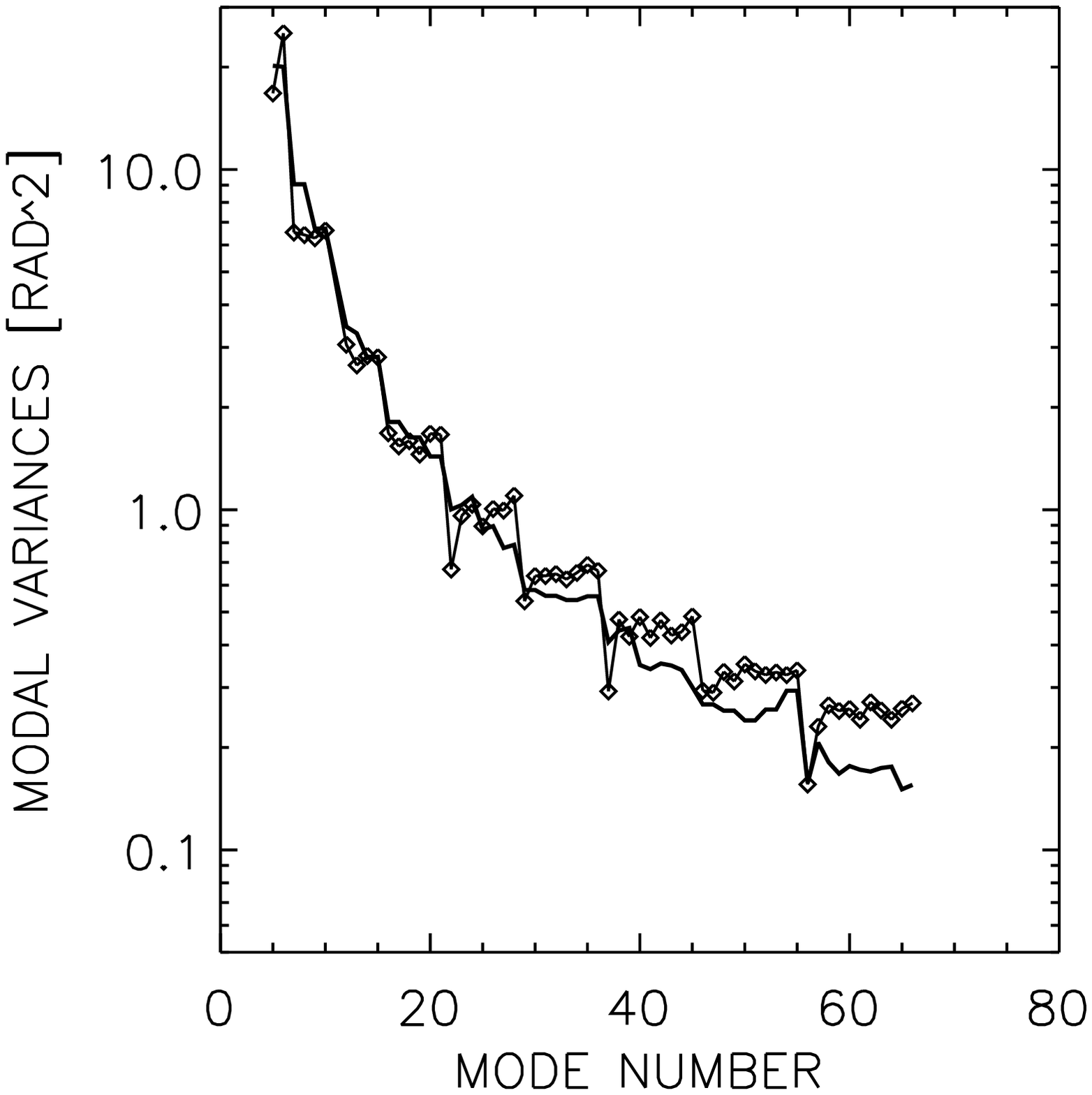}
   \includegraphics[height=8cm]{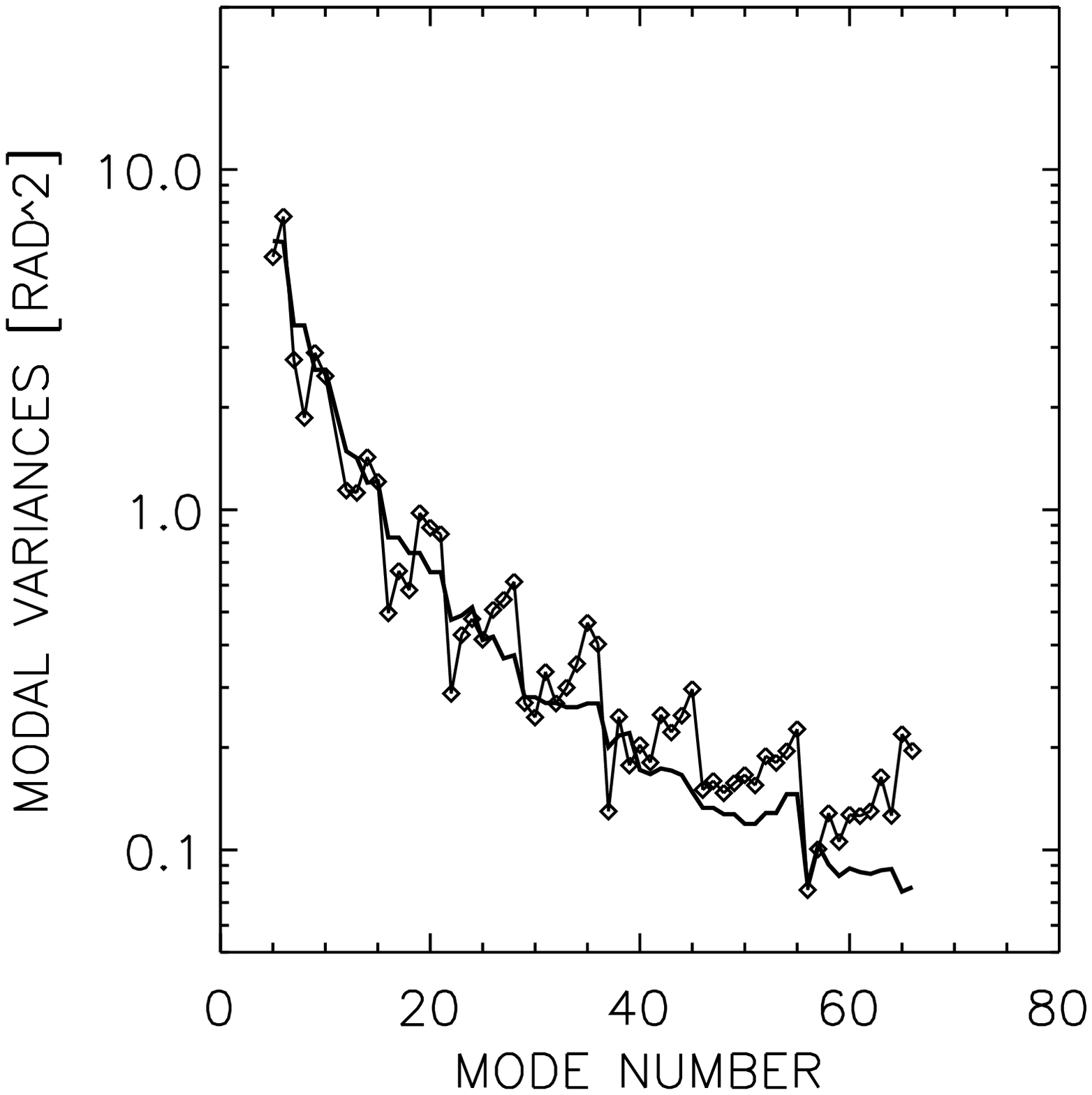}
   \caption{DM modal commands variances (diamonds) fit to the Von-Karman model (thick continuous line). Left: case 03/12/2009, seeing 0.93"; Right: case 25/12/2009, seeing 0.63".}
   \label{fig:fig02}
\end{figure}

Second, the modal variance distribution as a function of the mode index follows reasonably well the expected Von-Karman based distribution (see figure \ref{fig:fig02}) but if we look at the details of these distributions, there are cases (nights) where there was a strong departure with the VK-based model within the radial orders: one of the worse case is represented in the right of the figure, for the night 25/12/09. In this particular case (but also for other nights), within a given radial order, the departure with the model increases with the azimuthal order, as if an additional aberration, with some azimuthal periodicity, was added to the turbulent aberrations. This would be a first sign of the Gemini-North telescope secondary mirror print-through error an error identified after Altair commissioning. For other nights, this additional error does not appears in the DM commands (see figure,left, for the night 03/12/09). In short, there is, for certain nights, an additional non-turbulent based wavefront error which is not accounted for in our model. Knowing that an error on $r_0$ estimate has a direct impact on the fitting error structure function amplitude and the aliasing covariance matrix, this issue is very critical and need to be solved.

\subsection{PSF width excess and tilt jitter}

\begin{figure}[ht]
   \centering
   \includegraphics[height=7.8cm]{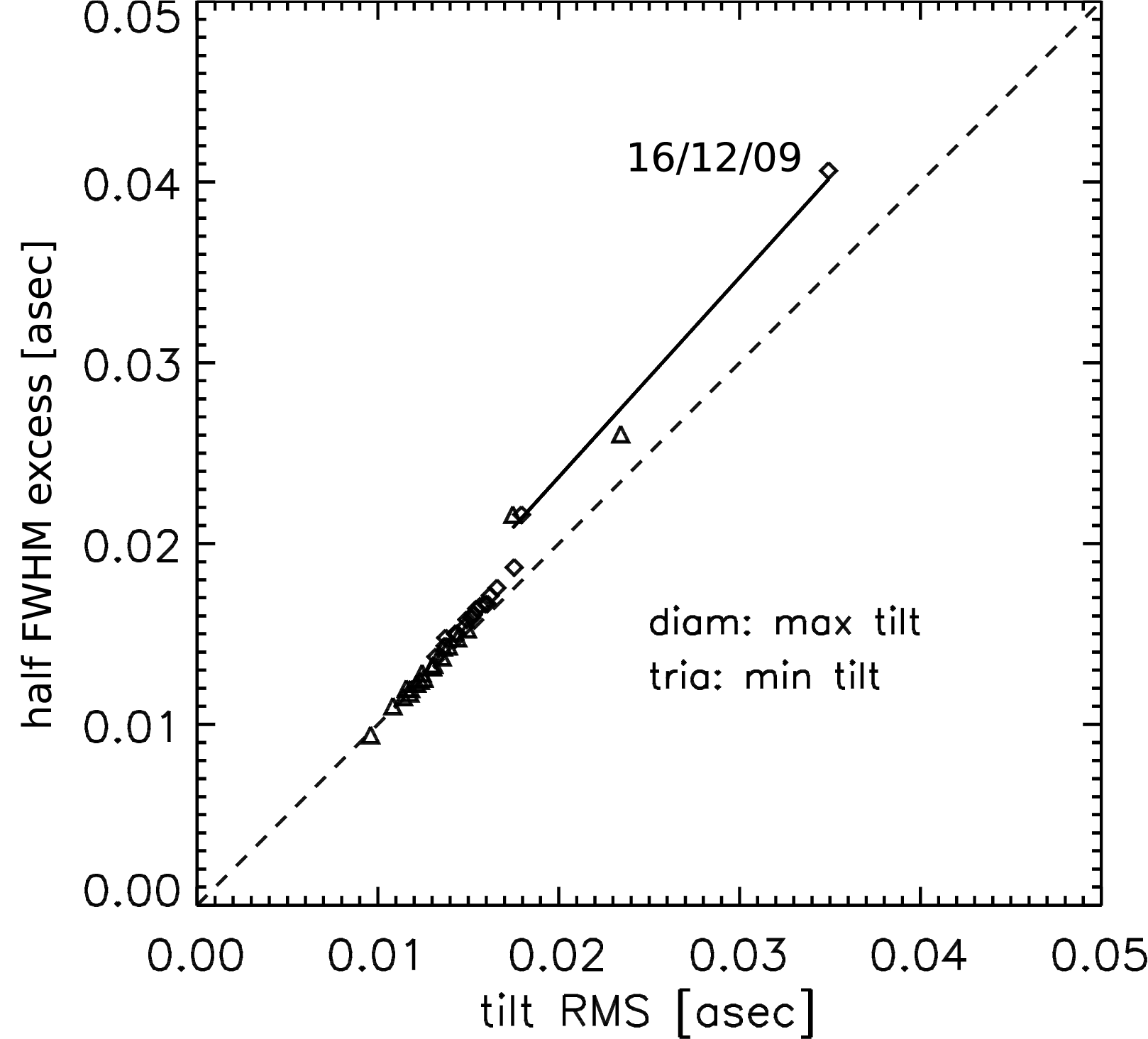}
   \includegraphics[height=7.8cm]{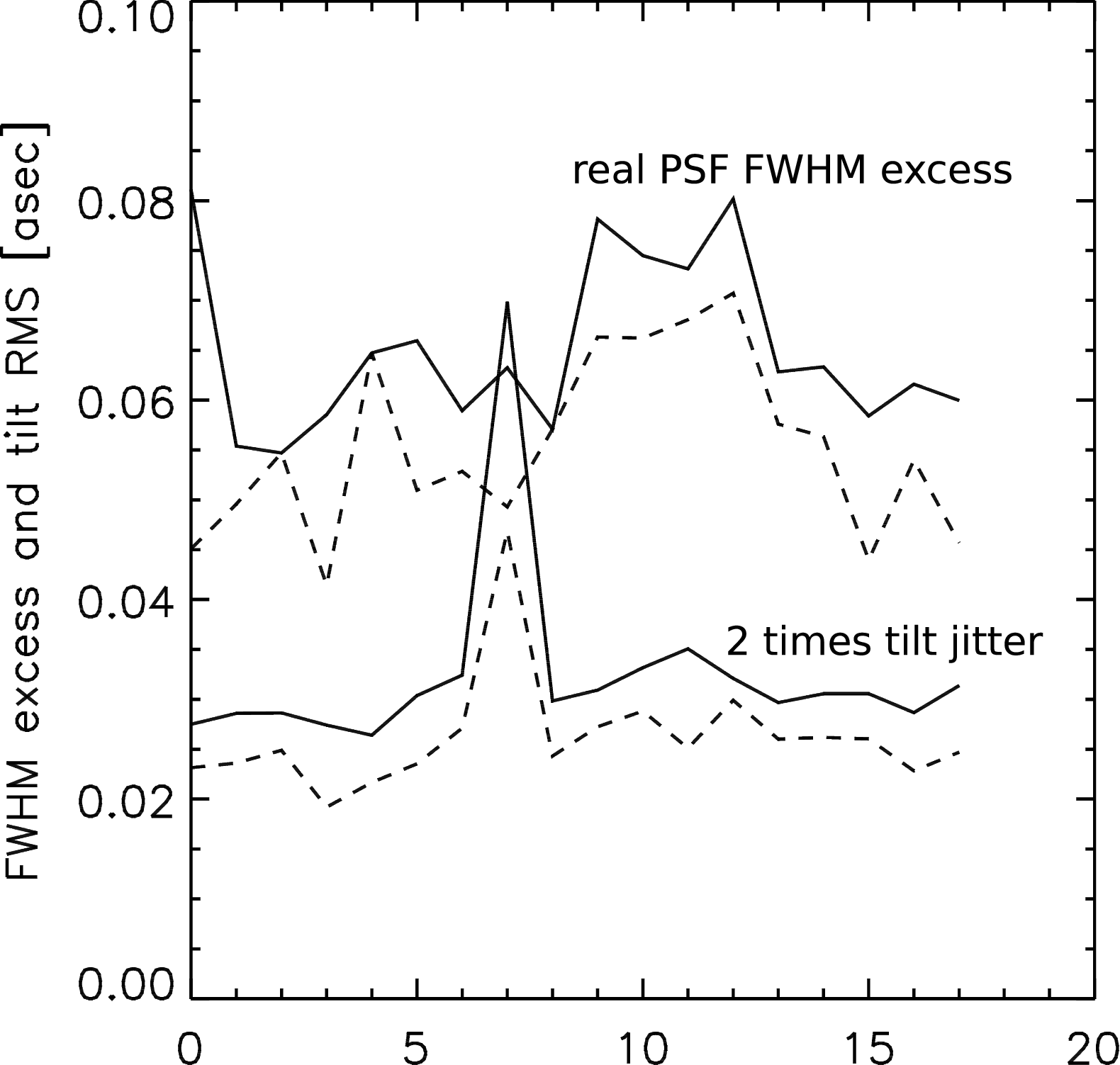}
   \caption{Left: half of the FWHM excess as compared with the tilt jitter RMS seen by the WFS. The tilt jitter and FWHH were measured along the maximal jitter and PSF core elongation. Right: for 18 out the 20 nights for which we had both the telemetry and the actual PSF, we show the FWHM excess expected from the WFS data (equal to twice the tilt jitter) and the FWHM excess measured on the real PSF.}
   \label{fig:fig03}
\end{figure}

What the WFS sees will be reconstructed in the PSF by the algorithm. One of the main, classical issue in AO control are vibrations. These can occur anywhere in the optical system before or after the WFS. Comparing the PSF width excess (with respect to the diffraction limit) on the reconstructed PSF with the width excess of the actual PSF can indicate the amount of non common path tilt jitter. Consider figure \ref{fig:fig03}. In the left, we show the FWHM excess (defined as the FWHM minus squared the diffraction limited FWHM) on the reconstructed PSF as compared to the tilt jitter RMS seen by the WFS. Note that we are showing the FWHM excess along the PSF maximal elongation, not in x and y, same for the tilt jitter. We see that there is an excellent agreement between the tilt jitter seen by the WFS and the reconstructed PSF FWHM excess. Actually the FWHM excess is slightly above the tilt jitter and this is due to the fact that higher order aberrations also contribute to the PSF core spread. The highest FWHM excess at 0.041" is for the night of 16/12/09, when the tilt jitter was particularly strong, for unknown reasons.

The right figure shows in the same plot the tilt jitter (min and max, again measured along and perpendicularly to the main jitter elongation) and the FWHM excess on the \textit{real} PSF. Apart for the night 16/12/09 (tag number 7) we see that there is basically no correspondence between what the WFS sees and the real PSF FWHM excess. The real PSF experiences a FWHM excess which is about twice what we would expect from the WFS measurement, i.e. about 60 masec instead of 30 masec. We interpret this as a demonstration that there are non common path vibrations in the system, after the WFS beamsplitter. A WFS measurement as close as possible to the science instrument would be required to tell. For now, it is impossible to go ahead with the PSF-R project with such a large jitter issue. If at least this jitter excess could be estimated, it would not be a difficulty to add it to the reconstructed PSF (convolution with a Gaussian).

\subsection{Reconstructed PSF, Strehl, wings structure}

In order to give us a sense of what would be the performance of Altair if the system was not suffering from hidden aberrations, we computed the long exposure PSF using our AO modeling tool PAOLA, a Fourier model for closed loop, single NGS AO systems. In K-band (2.125 $\mu m$), it is found that for the median MK conditions and the current system setup (wind 20 m/s, seeing 0.5", outer scale 30 m, loop rate 1 kHz, loop lag 0.8 ms, other parameters of Altair on Gemini apply) the Strehl would be 75\%. Now, the median Strehl in the reconstructed PSF, specifically ignoring the case 16/12/09, is in the range 45-67\%, with a median around 60\%. So the WFS apparently sees more aberrations than what it would seen for a perfect system, something in the range 110 nm to 240 nm from our Strehl estimation above. Now, this can be expected because there are always aberrations of complex origin that are not accounted for in a Fourier model, for instance calibrations errors, misalignments and so on. So from this exercise we simply can conclude that from the point of view of what the WFS sees, Altair performs relatively close to what we should expect for a perfect system.

\begin{figure}[t]
   \centering
   \includegraphics[height=7.8cm]{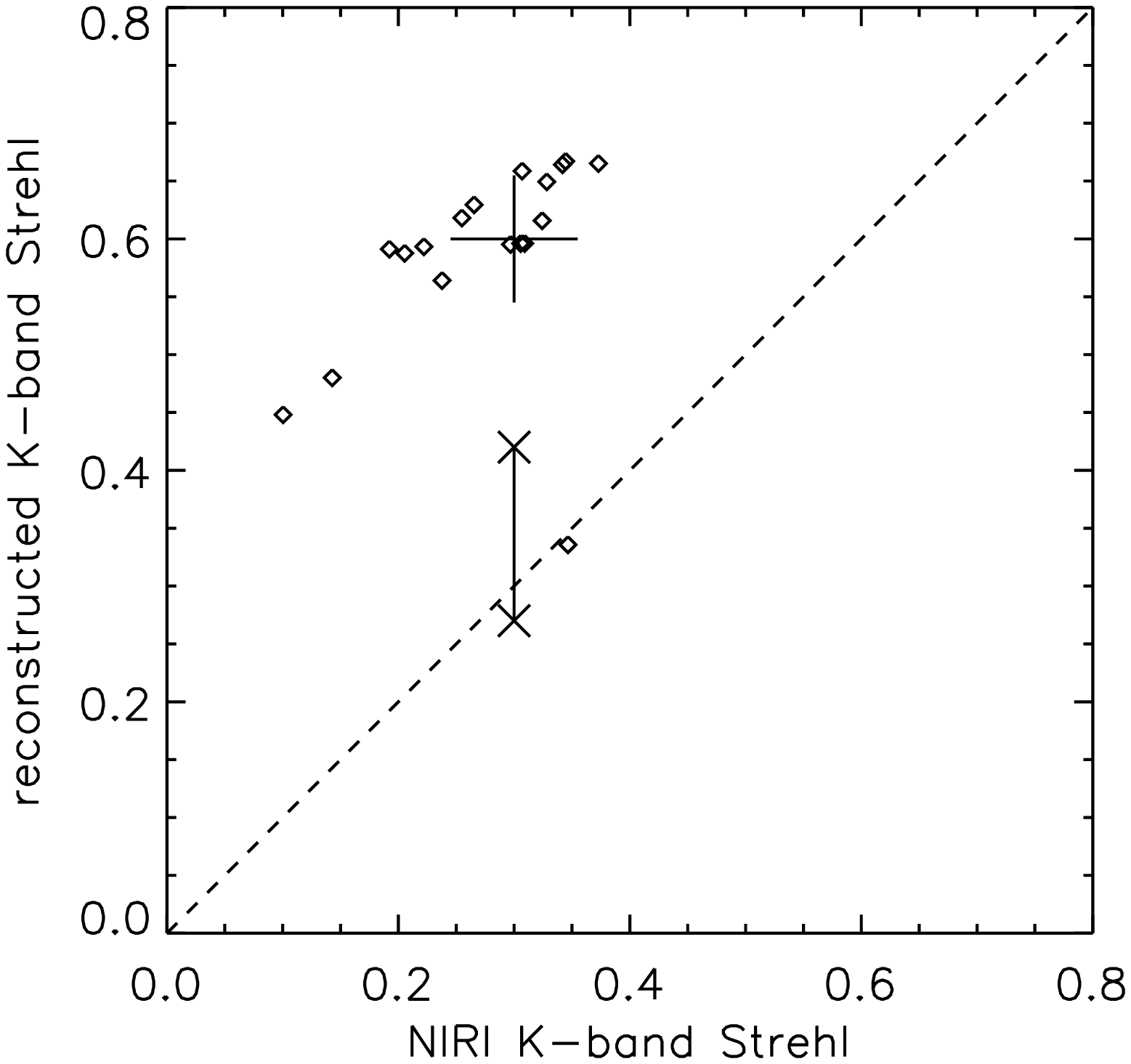}
   \includegraphics[height=7.8cm]{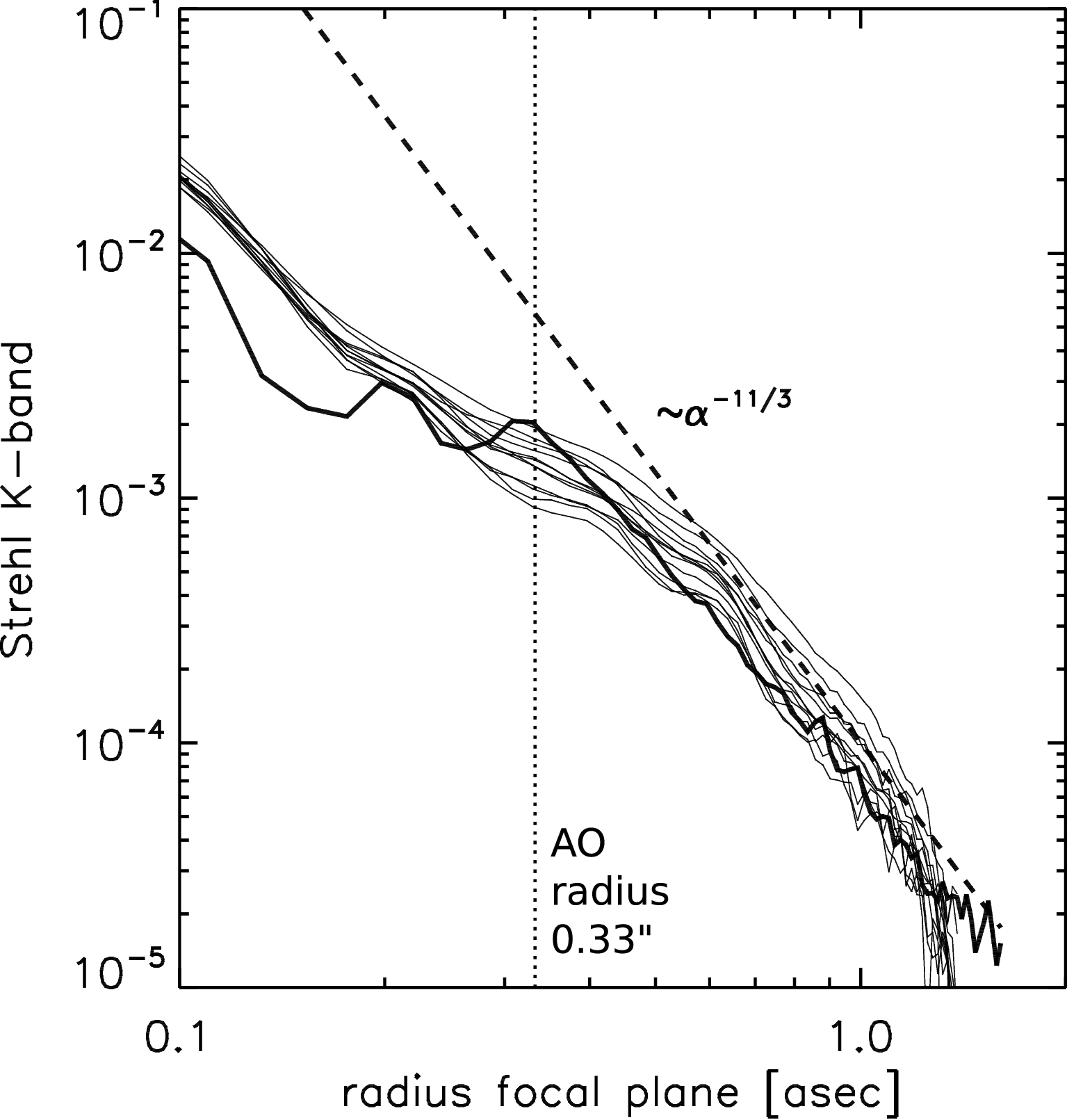}
   \caption{Left: Reconstructed versus measured K-band Strehl for December 2009. The large cross indicates the median values for both Strehls. Top x-cross indicates the median Strehl after correction with a fully correlated tilt-jitter Strehl (see text) and the bottom x-cross after correction with a totally uncorrelated tilt-jitter Strehl. Right: profile of the circularly averaged real PSF wings as compared to the expected -11/3 power law (in focal plane coordinate). The thin lines represents a few selected real PSF profiles, and the thicker line shows the x-profile of the reconstructed PSF for the night 08/12/09. The AO correction radius is indicated with the dotted vertical line.}
   \label{fig:fig04}
\end{figure}

But let us have a look now at figure \ref{fig:fig04}. It shows the reconstructed Strehl (K-band) as compared to the Strehl measured on the real, focal plane (NIRI) PSF. Basically, the reconstructed Strehl (median 60\%) is about two times larger than the measured Strehl (median 30\%). What can be the origin of such a discrepancy ? Actually we already know that the tilt jitter seen at the focal plane is about twice the tilt seen by the WFS. This excess of tilt jitter means that the energy in the PSF core is necessarily spread over a larger area than for the reconstructed PSF, therefore the real PSF Strehl has to be lower than the reconstructed one.

So the question is: can this jitter excess account for the Strehl loss ? Let us consider two extreme cases. First, assume that the additional unknown jitter is uncorrelated with the jitter seen by the WFS. As the focal plane jitter is about 30 masec and the WFS jitter is 15 masec, this means that there would be an additional jitter of 26 masec after the WFS (quadratic difference). In K-band, such a jitter would generate a very significant Strehl drop of 45\%. Multiplying the median reconstructed Strehl by 0.45, we find a post jitter Strehl of 27\%, a value in very good agreement with the real median Strehl (bottom cross in figure \ref{fig:fig04}). Second, let us consider that the additional jitter is totally correlated with the WFS jitter. In this case, we would have an excess of 15 masec, linearly added, which would correspond to a Strehl drop of 72\%, i.e. a post jitter Strehl of 43\%, compatible with the best real Strehl values (top cross in figure \ref{fig:fig04}). The answer to the question above is therefore yes, the additional tilt might explain part of the post-WFS Strehl loss. Not all, because there must be some room left for other errors, like instrument PSF and other high order non common path errors.

Reproducing the structure of the PSF wings is an other important aspect of PSF-R, as it has a direct impact on the integrated energy performance, of particular importance in aperture photometry or spectroscopy data reduction. It is more difficult though, because it requires an accurate knowledge of not only the residual DM modes, but also the PSF associated with all the non-sensed aberrations of the system. V\'eran's method to get the telescope PSF, as explained above, works as long as the telescope and other hidden aberrations are static, but if there are non turbulent dynamic aberrations, there is a risk that these would not be reproduced between the science AO run and the calibration run. The structure of the reconstructed PSF is shown in figure \ref{fig:fig04}, right, and we found out that as expected the PSF wings above the AO correction radius (here $\lambda/(2\,\text{pitch})=0.33"$) follows the trend in $x^{-11/3}$ of a seeing limited PSF, given by the Kolmogorov spatial power spectrum of the un-corrected phase.

What is interesting though is that the real PSF wings -- after circular averaging -- show the same trend, beyond the AO correction radius. This tends to demonstrate that there are no significant radial structure additional error in the telescope optics. Of course a very good candidate for this type of error would be the residual mirror polishing error, but we know from a study from Ren\'e Racine (personal communication) that such an error amounts to no more than about 7 nm RMS, and is therefore totally burden by the residual turbulent (fitting error) aberration in our case. So, the overall radial profile of our reconstructed PSF seems to fit relatively nicely the real PSF wings average structure, above the AO correction radius.

What about the detailed structure, though ? We show in the figure \ref{fig:fig05} the wings details (linear scaling) for one of the nights where the structure in the wings were particularly apparent. It is important to remember here that the exposure time was about 1 minute, therefore we should expect the optical turbulence speckles to have been smoothed out. Now, speckles are still present and dominate the PSF wings. We speculate therefore that these speckles originate from an other source than the atmosphere, and the telescope M2 support structure print-through wavefront error -- identified in extra focal images and inspection of the residual high order truth WFS of Altair during post commissioning tests (V\'eran, Rigaut, personal communication) -- is thought to be at the
\begin{floatingfigure}{80mm}
\begin{center}
\includegraphics[width=75mm]{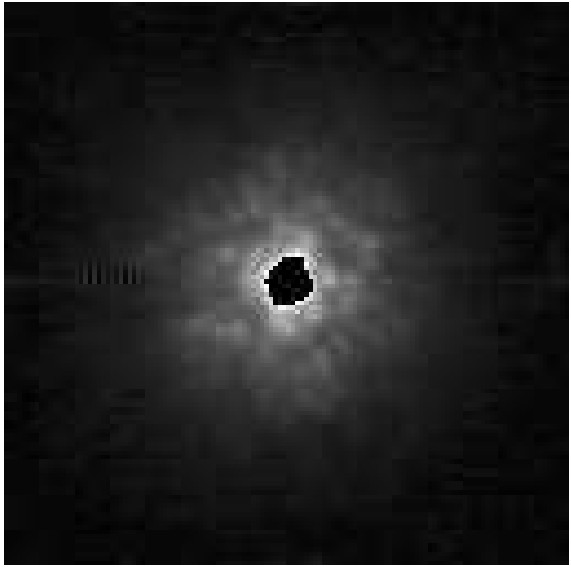}
\end{center}
\caption{\label{fig:fig05} Wings of the real PSF (22/12/09). Exposure time 1 minute. $\sqrt{\text{PSF}}$ scaling, the PSF is cut at 5\% of its maximal peak value. Field of view is 2.8", pixel size 0.022". K-band, NIRI image.}
\end{floatingfigure}
origin of these structures. Indeed, this M2 error affects mainly the spatial frequencies above the AO cutoff spatial frequency, therefore it is (a) not corrected by the system, (2) aliased by the WFS into the low spatial frequency domain (i.e. affects the PSF structure inside the correction radius).

A quantitative estimate of the impact of this error on the AO performance has been done by V\'eran, and is expected to be on the order of 100 nm at high order and in the range 30 to 150 nm at low order (aliasing component), which is equivalent to a Strehl drop of about 70\% in K-band. These numbers are only indicative though, as the behavior/impact of the M2 print-through error highly depends on the registration of M2 with Altair's WFS lenslet array, which varies from an exposure to another, depending on the field rotation compensator at the Cassegrain focus. And M2 was replaced from the time of V\'eran ad Rigaut's study, but still shows the same print-through error. This being said, this error has a systematic and predictive behavior, and the question remain to what extent it would be possible to model it and include it in our PSF-R model, or to compensate for it while running the AO loop.

\section{THE FUTURE OF PSF-R: A DISCUSSION, AND NEW IDEAS}

\subsection*{PSF-R algorithm and AO system design}

One of the first comment we can make here is that our tentative on Altair at least has clearly demonstrated the interest of PSF-R for system performance diagnostic. Indeed, comparing the reconstructed PSF with the actual one has allowed us to show that the system suffers from an excess of tilt jitter after the WFS beamsplitter, and that there is some complicated non-Kolmogorov wavefront error affecting the beam before the DM - as we can see in the DM modal commands RMS values. So, PSF-R fulfills one of its designated role here: system diagnostic.

AO astronomers, of course, have other needs, which can be sorted in two categories:
\begin{enumerate}
\item they need real time basic PSF metrics estimates (Strehl, FWHM, integrated energy) but also indicators of the current state of optical turbulence (how strong and how fast is it) in order to optimize the observation on-the-fly, for instance optimizing the imager exposure time, or optimizing telescope time use by quickly pointing to another science target/program, depending on the seeing conditions
\item the PSF associated with the AO run is the other thing they need, but there is no need for real time here as AO data reduction is done later. It is important that the AO astronomer, when leaving the telescope with the AO data, also gets the associated PSF, because PSF-R codes firstly need a lot of inputs that can take lots of memory space (can be Terabytes), and secondly if a telemetry data was missing or corrupted at the time of the data gathering, post PSF-R would not work. We think it is therefore mandatory to have the PSF-R algorithm running \textit{almost} in real time, or at least immediately after the completion of each AO run.
\end{enumerate}

Fulfilling the first requirement -- instantaneous metrics -- is certainly possible but would require an other approach than V\'eran's, which relies on "long exposure" statistical quantities. Optical turbulence metrics can be estimated from a large number of sources: from the instantaneous local \cns\ and wind velocity profiles, and this would be the ideal case but such profiles are not always available at the exact location of the telescope, or from AO telemetry data analysis (for instance the WFS measurements temporal power spectrum might be used to retrieve an estimate of the average wind velocity, and the DM commands might be used to estimate the overall seeing, actually as in V\'eran's approach, but possibly with less accuracy as we would use much less samples than in the long exposure approach). There is clearly some work to do here in developing these metrics estimates.

The second requirement, classical PSF-R, as we have seen, might work, buy only if we have a good model \textit{and} measurement of the additional wavefront static AND dynamic errors generated by the system telescope + AO + instrument. In the Gold Age of AO, AO system were devised for existing telescopes -- some were already respectable old machines -- and when switching on the system for the first time, AO engineers were basically discovering all the hidden aberrations of the telescope system that were not always accounted for in the AO system design; PSF-R was tried for these systems, sometime with success, because the "working environment/conditions" of the telescope and AO system were good (V\'eran success on PUEO/CFHT) but most of the time with failure, essentially for the reasons indicated in the introduction.

Today, telescope design has to met requirements specified by the client AO system, in order to be sure that AO performance will not be limited by the telescope. Doing PSF-R for such systems might be easier than on previous machines, as the source of hidden aberrations are potentially less numerous, and possibly, non-Kolmogorov static/dynamic aberrations can be known in advance and V\'eran's model updated to account for those. There is a big risk, though: if the PSF-R algorithm design comes AFTER telescope and AO system design and manufacturing, it might be difficult or impossible to get access to reliable estimates of these non-Kolmogorov wavefront errors. In these conditions, PSF-R would need to rely on best guesses, possibly empirical modeling, as we are doing (will do) for Altair on Gemini, and PSF-R might even be a failure. To avoid such a dramatic issue -- not knowing the PSF of a segmented giant telescope costing a billion euros/dollars \textbf{is} dramatic -- we believe that \textbf{PSF reconstruction requirements have to be taken into account at the level of the design of the AO system}, in order to make sure that telemetry for all (significant) sources of errors will be available for PSF-R. It might even be that other data than telemetry is required too, for instance sky (focal plane) data, which might require some specific hardware/software to be implemented in the AO system design, so again, it is critical that the PSF-R algorithm design be done in common with the AO system (and possibly instrument and telescope) design.

\subsection*{Poor's man PSF-R}

Now there are cases where the ideal situation of full data access discussed above is not possible, or not yet implemented, or simply not realistic for whatever reasons. In these conditions, an option would be to make use of analytic AO PSF models (end-to-end Monte-Carlo models would be much too slow) where the usual model's inputs (turbulence, AO system) would be taken from estimates of the AO run conditions (seeing monitor, AO loop settings). Such analytic models have been demonstrated to produce PSF estimates in good agreement with end-to-end Monte-Carlo codes, and can be easily modified to take into account non-Kolmogorov wavefront errors (telescope, vibrations, ...) as long as an equivalent OTF filter can be devised. One of such tool is PAOLA, cited above. There is also a GLAO mode in PAOLA that was tested against a very realistic end-to-end Monte-Carlo code (OCTOPUS, Miska Le Louarn, ESO) and produced very good estimate of the long exposure PSF structure (Strehl, wings structure, everything). Neichel et al. \cite{neichel:08} have proposed a model for tomography and multi DM correction based on the same analytic (Fourier) approach that might be also a very good start for "cheap" PSF-R in the case of MCAO/LTAO systems. PAOLA is a good example of a code that is already available and could be tested immediately on real sky data. Looking and understanding the differences between the analytically reconstructed PSF and the actual on-sky PSF, and implementing a model for these additional errors would allow to have at one's disposal a fast and easy to use PSF-R tool.

\subsection*{PSF-R using focal plane (sky) data}

Let us discuss now the case of PSF-R for multi NGS/LGS and mono/multi DMs (GLAO, LTAO, MCAO, MOAO). A very important request for PSF-R on such systems is the ability to predict the PSF all over the field-of-view. Such systems, though, are significantly more complex than single, on-axis NGS AO systems, and the residual wavefront error has multiple origins: for instance the wavefront error budget for Keck II AO has 7 components in NGS mode, 12 in LGS mode; about 16 for ATLAS, an LTAO instrument for the E-ELT\footnote{European Extremely Large Telescope; Fusco, T. et al. this conference}, and more than 20 for NFIRAOS, the first light MCAO instrument for TMT\footnote{Thirty Meter Telescope; Gilles, L. et al. this conference}.

The question therefore is: are we going to build/extend V'eran's like model for all these errors, knowing on top of this that cross talks between error terms are even often ignored in errors budgets ? Of course, such sophisticated models have to be developed, but it might be wise to seek alternate solutions if the telemetry needed to calibrate such models, or the models themselves, turns out to be too cumbersome. Besides, it is doubtful that telescope time on these costly multi-GS systems can be devoted to the observation of single natural guide stars to retrieve the telescope + instrument OTF, using the procedure proposed by V\'eran for single NGS systems.

An alternative to V\'eran's method, or a complement to this method, can be to make use of the NGS and LGS focal plane images themselves. Normally, the NGS and LGS light beams are captured at the entrance of the AO system (using either pick-up arms or beam splitters) and sent to the different WFS of the system. If part of the light of these GS would be left to propagate until the instrument input focal plane, or at least as close as possible to this plane, and retrieved using appropriate detectors (WFS or imager) then it would be possible to get, for different directions \textit{inside the science field}, a measure of the actual AO corrected PSF that includes all errors, \textit{including the telescope one}.

To say things differently: there are GS within or close to the science field and the images of these GS can certainly be used as calibration PSF to estimate the system performance across the science field of view. Inter/extrapolating these calibration PSF over the field would be done by using a generalization of Britton's method \cite{britton:06}, where off-axis PSF are modeled from measured on-axis PSF using a \cns-calibrated angular anisoplanatism structure function model. Such a generalized angular anisoplanatic structure function can certainly be built for multi-GS (we haven't tried) using, for instance, the analytical approach developed by Neichel et al. (indicated above).

In the ideal case, no telemetry data would be required, as the knowledge of the \cns\ profile and the system geometry would be sufficient to build such a structure function. Practically, though, owing to the system's complexity, it is very probable that we would need telemetry data to complement the data gathered from the sky, and/or to retrieve the apparent \cns\ profile if no \cns\ profiler (not close or fast enough) is available for the telescope. In fact, it might well be that the solution for reliable (rich's man) PSF-R on multi-GS/DM AO systems will require both full telemetry and PSF calibrator across the field-of-view. Such system are currently in study \& development (some of them are near first light) so we believe now is the good time to study these different PSF-R options.

\section{CONCLUSIONS}

In this article we describe the first on-sky test of a point spread function reconstruction (PSF-R) algorithm for Altair, the Gemini-North telescope adaptive optics system, in natural guide star mode. It is found that the method does work, but suffers from hidden telescope and non common path wavefront errors (vibrations, M2 support structure print through) that have a strong impact on the algorithm performance. Either we need to update our PSF-R algorithm to take these errors into account, or these errors have to be eliminated at their source, before an automatic PSF-R tool can be made available for Altair's users. This being said, PSF-R has demonstrated its usefulness for system performance diagnostic, as by comparing the PSF as it is seen from the WFS and the actual on-sky PSF, one can estimate qualitatively and quantitatively the hidden (non common path) aberrations of the system.

Based on this preliminary experience, we stress the importance of including PSF-R requirements at the core of future AO systems design, particularly for complex multi guide stars \& multi deformable mirror systems. We propose also alternative PSF-R techniques as back-up solutions in cases where telemetry is not or partially available, or when telemetry needs to be complemented with focal plane (sky) PSF data.

\acknowledgments
 
L.J. would like to thank Eline Tolstoy from Kapteyn Astronomical Institute, (Netherlands) and Peter Wizinowich, W. M. Keck Observatory, Hawaii (United States) for their generous support during this study. The authors would like to thank also Jean-Pierre V\'eran, Fran\c{c}ois Rigaut and Benoit Neichel for many fruitful discussions concerning Altair telemetry and the Gemini-North telescope optics.

The Gemini Observatory is operated by the Association of Universities for Research in Astronomy, Inc., under a cooperative agreement with the NSF on behalf of the Gemini partnership: the National Science Foundation (United States), the Science and Technology Facilities Council (United Kingdom), the National Research Council (Canada), CONICYT (Chile), the Australian  Research Council (Australia), MinistŽrio da Cincia e Tecnologia (Brazil), and Ministerio de Ciencia, Tecnolog'a e Innovaci—n Productiva (Argentina).


\begin{thebibliography}{1}

\bibitem{veran:97}
J.-P. V{\' e}ran, F.~Rigaut, H.~Ma{\^ i}tre, and D.~Rouan, ``{Estimation of the
  Adaptive Optics Long-exposure Point Spread Function using Control Loop
  Data},'' {\em {Journal of the Optical Society of America A}}~{\bf 14},
  pp.~3057--3069, 1997.

\bibitem{jolissaint:04}
L.~Jolissaint, J.-P. V\'eran, and J.~Marino, ``{OPERA, an automatic PSF
  reconstruction software for Shack-Hartmann AO systems: application to
  Altair},'' in {\em Astronomical Telescopes and Instrumentation},  {\em SPIE
  Proceedings} {\bf 5490}, 2004.

\bibitem{lemignant:08}
D.~L. Mignant, R.~C. Flicker, M.~C. Britton, J.-P. V\'{e}ran, M.~P. Fitzgerald,
  D.~T. Gavel, L.~Jolissaint, M.~A. van Dam, and E.~M. Johansson, ``{AO PSF}
  reconstruction: a review of the quests,'' {\em SPIE Proceedings} {\bf 7015},
  2008.

\bibitem{jolissaint:06}
L.~Jolissaint, J.-P. V\'eran, and R.~Conan, ``{Analytical Modelling of Adaptive
  Optics: Foundations of the Phase Spatial Power Spectrum Approach},'' {\em
  {Journal of the Optical Society of America A}}~{\bf 23}(2), pp.~382--394,
  2006.

\bibitem{neichel:08}
B.~{Neichel}, T.~{Fusco}, and J.-M. {Conan}, ``{Tomographic reconstruction for
  wide-field adaptive optics systems: Fourier domain analysis and fundamental
  limitations},'' {\em Journal of the Optical Society of America A}~{\bf 26},
  p.~219, 2008.

\bibitem{britton:06}
M.~Britton, ``{The Anisoplanatic Point-Spread Function in Adaptive Optics},''
  {\em {Publications of the Astronomical Society of the Pacific}}~{\bf 118},
  p.~885, 2006.

\end{thebibliography}
\end{document}